\documentclass[12pt,a4paper,reqno]{amsart}
\usepackage{xypic,a4wide,amsmath,amssymb,amsthm}
\usepackage[cp1251]{inputenc}
\usepackage{bm}
\usepackage{amssymb}
\usepackage{amsmath}

\usepackage[dvips]{graphicx}
\usepackage{array}

\title[\ ]{Scattering and Orbiting in Reissner-Nordstrom black hole spacetime}

\author{Sadia Nazir$^1$}
\address{
Department of Physics, The University of Lahore, Pakistan} \email{sadia.nazir@phys.uol.edu.pk} \urladdr{}
\title[\ ]{Scattering and Orbiting in Reissner-Nordstrom black hole spacetime}

\author{ M.A.Faridi$^2$}
\address{
Centre For High Energy Physics, University of the Punjab, Lahore(54590), Pakistan} \email{ayubfaridi.chep@pu.edu.pk} \urladdr{}

\begin{document}

\maketitle

\begin{abstract}
 By using semi-analytical method,we investigate the transmission and reflection coefficient and absorption and scattering cross section of spin 0,1,2 particles which are massless from a charged nonrotating black holes. Here orbiting scattering can be characterized of massless particles from Reissner-Nordstrom black hole at a critical angular momentum. We examine the effects upon reflection,transmission,scattering and absorption cross section due to the black hole charge. We find that with the increase of charge the absorption cross section decreased in spin-2 particle while in case of spin 0 particle it is reversed and in case of spin 1 the effect is negligible due to charge.
\end{abstract}
\section{Introduction}

For understanding of the atomic nuclei one of the most interesting part is scattering. In modern physics black holes play significant role and is being distinguished by their angular momentum,mass and charge \cite{01,02}. Ford and Wheeler developed a connection between classical and quantum mechanical counterparts in 1959 \cite{03}. For the early study of the black holes analytical methods were used but with the development numerical techniques attains great attention in which computers (simulation) are used for the study of the absorption and scattering problems of the black holes. With the discovery of gravitational waves from the smashing of black holes, LIGO has laid down the roots of universe. Hopefully in near future black hole study will receive more importance for understanding of the universe.
 \newline
 Up till now scattering by Schwarzschild black holes for massless,massive scalar fields and waves have got more attention as compared to the others \cite{04}-\cite{11} in static black holes space times. In literature considerable work is present for Reissner-Nordstrom metric for massless \cite{12} and massive scalar fields \cite{13} ,electromagnetic \cite{14}-\cite{16} and gravitational wave \cite{17,18}.
 Using higher-dimension some work can also be found for Reissner-Nordstrom metric \cite{19,20,21}. The conversion coefficient of electromagnetic to gravitational waves were computed by Luis et al \cite{22}.
 Similar structure is observed for Bardeen spacetime as Reissner-Nordstrom black hole. These spacetimes have two horizons. Spiralling trajectories would be observed for below and above the critical values of the angular momentum and at $l_{c}$ one could expect the limit of an unstable orbit being circular. The condition corresponding to an unstable orbit for an incident particle having energy $E$  and effective potential $V_{f}$ can be $V_{f}(r_{c},l_{c})=E$.
 \newline
 Most of the work in literature for Reissner-Nordstrom black holes was done numerically in which spin part is missing. Although the present work is done by using semiclassical approximation but its includes different spin massless scattering particles. We evaluate reflection and transmission coefficients at a different values of angular momentum and spin for the extreme case of Reissner-Nordstrom metric. We also calculate scattering and absorption cross section. We observe that the critical angular momentum is nothing but it significantly corresponds to an unstable circular orbit. By using almost same formalism D.Batic et al \cite{23} evaluated the reflection coefficient for the Schwarzschild black hole case.
\newline
The line element of Reissner-Nordstrom spacetime is as follow
\begin{equation}\label{eq:1}
\left\{
 ds^2=y(r)dt^2-y(r)^{-1}dr^2-r^2(d\theta^2+\sin^2\theta d\phi^2)
\right.
 \end{equation}
 where
\[  y(r)=(1-\frac{r_{+}}{r})(1-\frac{r_{-}}{r})  \]
and
\[ r_{\pm}= M\pm\sqrt{M^2-Q_{N}^2} \]
By using \[ c=G=1\] i.e natural units and metric signature $(+ - - -)$.
We exhibit results for extreme case where $M=|Q_{N}|$ in Reissner-Nordstrom metric.
\\Where
$y(r)=1-\frac{2M}{r}+\frac{Q_{N}^2}{r^2}$
\\By substituting the metric into Klein-Gordon equation,
$ \square\xi=0$
\\Using separation ansatz $U(t,r,\theta,\phi)= \frac{\xi(r)}{r} Y_{lm}(\theta,\phi)e^{-\iota\omega t}$,we get following radial equation
$$y(r)\frac{d}{dr}(y(r)\frac{d\xi}{dr}+[\omega^2-v_{f}(r)]\xi=0$$
\begin{equation}\label{eq:2}
\left\{
v_{f}(r)=y(\frac{y'}{r}+\frac{l_{r}(l_{r}+1)}{r^2})
\right.
\end{equation}
\begin{figure}[htb]
\centering
\includegraphics[width=10cm]{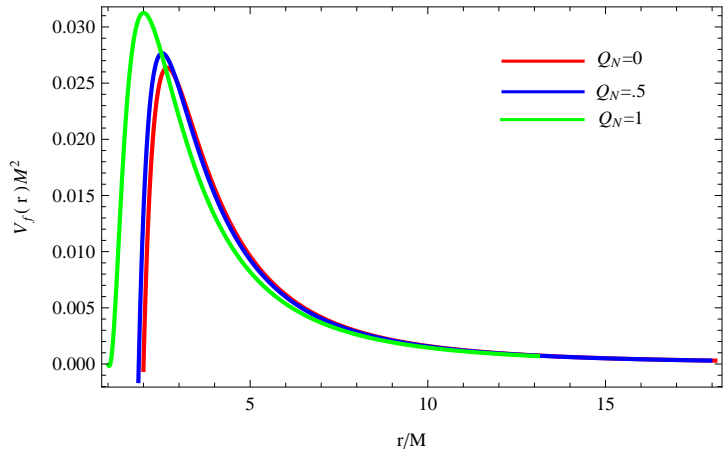}\\
\caption{Effective potential for fixed value of $l_{r}$ and different values of $Q_{N}$ of Reissner-Nordstrom black hole.}
\label{fig-smp1}
\end{figure}
In fig.1,effective potential with a fixed value of $l_{r}$ i.e $0$ is plotted for three different values of $Q_{N}$ as a function of $r$. As we can see that at the  infinity or event horizon the effective potential goes to zero and it is maximum as the charge increased.
$$\frac{d}{dr_{*}}=y(r)\frac{d}{dr}$$
or in integral form
$$x_{*}=r+\frac{r_{+}^2}{r_{+}-r_{-}}\ln\mid\frac{r}{r_{+}}-1\mid-\frac{r_{-}^2}{r_{+}-r_{-}}\ln\mid\frac{r}{r_{-}}-1\mid+c$$
Where $c$ is an integration constant and we set $ c = 0$ making no influence of our numerical results.
\begin{figure}[htb]
\centering
\includegraphics[width=10cm]{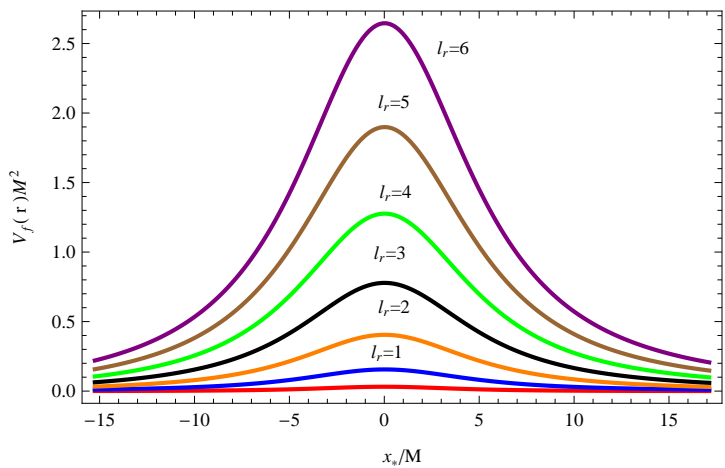}\\
\caption{Effective potential for extreme Reissner-Nordstrom black hole for different values of $l_{r}$.}
\label{fig-smp2}
\end{figure}
Fig.2 represents the effective potential, $V_{f}$ for extreme Reissner-Nordstrom black as a function of tortoise coordinate, $x_{*}$, for $|Q_{N}|=M$ for different choices of angular momentum,
$l_{r}$. As an evident maximum of the potential observed with the increase of $l_{r}$.
\begin{equation}\label{eq:3}
\left\{
\frac{d^2}{dx_{*}^2}\xi(x_{*})+[\omega^2-v_{f}(x_{*})]\xi(x_{*})=0
\right.
\end{equation}
Eqn.(3)can also be obtained for different spins i.e 0,1,2
\begin{equation}\label{eq:4}
\left\{
v_{f}(r(x_{*}))=(1-\frac{2M}{r}+\frac{Q_{N}^2}{r^2})[(\frac{2M}{r^3}-\frac{2Q_{N}^2}{r^4})(1-s^2)+\frac{l_{r}(l_{r}+1)}{r^2}]
\right.
\end{equation}
This equation holds with $l_{r}\geq s$
\cite{23,24}
\\For asymptotic limits of eqn.(3),we obtain the solution
\begin{equation}\label{eq:5}
\left\{
\xi(r)\approx\{
\begin{split}
B^{tr}_{\omega l} e^{-\iota \omega x_{*}}       for          x_{*}\rightarrow -\infty(r\rightarrow r_{+})
\\B^{out}_{\omega l}e^{\iota \omega x_{*}} + B^{in}_{\omega l} e^{-\iota \omega x_{*}}         for x_{*}\rightarrow +\infty(r\rightarrow \infty)
\end{split}
\right.
\end{equation}
with satisfying the relation
\begin{equation}\label{eq:6}
\left\{
|B^{in}_{\omega l}|^2=|B^{out}_{\omega l}|^2+|B^{tr}_{\omega l}|^2
\right.
\end{equation}
\textbf{\Large{\\[.2ex]Critical Angular momentum and occurring of Unstable Orbits}}
\\
 At $r=2M$ with $s=1$ in eqn.(4)\cite{25}an unstable photon orbit is expected for Reissner-Nordstrom black hole scattering.
\\As $l_{r}(l_{r}+1)\rightarrow(l_{r}+\frac{1}{2})^2$ in semiclassical prescription,we consider $v_{f}(2M)=\omega^2$,the critical value of $l_{r}$ (angular momentum) is given easily as
$l_{c}=4\omega r_{m}-\frac{1}{2}$    where  $r_{m}=M$.
By using same way too the occurrence of circular orbits due to critical $l_{c}$ of the scattering of spin $0$ and $2$ particles could be find. Consider $v_{f}(r)$ at $r=2M$ in eqn.(4)
$$l_{c}^{wkb}=\sqrt{16 \omega^2r_{m}^2-(1-\frac{Q_{N}^2}{2})(1-s^2)}-\frac{1}{2}$$
By using semiclassical approximation the critical values of the angular momentum
$$l_{c}^{qm}=\sqrt{16 \omega^2r_{m}^2-(1-\frac{Q_{N}^2}{2})(1-s^2)+\frac{1}{4}}-\frac{1}{2}$$
$v_{m}=v_{f}(r=2M)$, its values are different for different choices of spin,charge and angular momentum.
For$s=0,1,2$,we obtain
$$r_{m}^2v_{m}=\frac{1}{16}[l_{r}(l_{r}+1)+(1-\frac{Q_{N}^2}{2})],             s=0$$
$$r_{m}^2v_{m}=\frac{1}{16}[l_{r}(l_{r}+1)], s=1$$
$$r_{m}^2v_{m}=\frac{1}{16}[l_{r}(l_{r}+1)+\frac{3}{2}Q_{N}^2-3],s=2$$

\begin{table}[htb]
\caption{Critical values of angular momentum at $Q_{N}=.2$ are obtained by using $v_{f}$ at $r=2M$. The values of $l_{c}^{WKB}$ are present outside the bracket while inside is for $l_{c}^{QM}$.}
\label{tbl-smp1}
\vspace{2ex}
\begin{center}
\begin{tabular}{|c|c|c|c|}
\hline
 $\omega r_{m}$   & $s=0$    & $s=1$ & $s=2$\strut\\
  \hline
  \hline
  .5 &         1.237[1.308]          &       1.5[1.561]     &     2.134[2.181]\strut\\
  1 &         3.376[3.408]           &       3.5[3.531]     &     3.582[3.881] \strut\\
  1.5&         5.418[5.438]           &       5.5[5.521]   &     5.740[5.760] \strut\\
  2&         7.439[7.454]           &       7.5[7.515]      &     7.682[7.697] \strut\\
  \hline

\end{tabular}
\end{center}
\end{table}
Our potential is similar to eqn.(14) in Ref.\cite{23} by using $a=3.4$ and $b=.6$ and one can check by plotting it. Due to the similarities between these potentials and using analytical results from \cite{24} with $k^2=\omega^2$,$\alpha=\frac{1}{a r_{m}}$ and $2mU_{0}=v_{m}$ transmission coefficients become
\begin{table}[htb]
\caption{Critical values of angular momentum at $Q_{N}=.4$ are obtained by using $v_{f}$ at $r=2M$. The values of $l_{c}^{WKB}$ are present outside the bracket while inside is for $l_{c}^{QM}$.}
\label{tbl-smp1}
\vspace{2ex}
\begin{center}
\begin{tabular}{|c|c|c|c|}
\hline
 $\omega r_{m}$   & $s=0$    & $s=1$ & $s=2$\strut\\
  \hline
  \hline
  .5 &         1.255[1.325]          &       1.5[1.561]     &     2.10[2.148]\strut\\
  1 &         3.383[3.415]           &       3.5[3.531]     &     3.831[3.860] \strut\\
  1.5&         5.423[5.443]           &       5.5[5.521]   &     5.726[5.746] \strut\\
  2&         7.442[7.458]           &       7.5[7.515]      &     7.671[7.686]\strut\\
  \hline

\end{tabular}
\end{center}
\end{table}
$$|T_{l_{r}}|^2=\frac{\sinh^2(\pi a \omega r_{m})}{\sinh^{2}(\pi a \omega r_{m})+\cos^{2}\frac{\pi}{2}\sqrt{4v_{m}a^2r_{m}^2-1}}$$
if  $4v_{m}a^2r_{m}^2>1$ and
$$|T_{l{r}}|^2=\frac{\sinh^2(\pi a \omega r_{m})}{\sinh^{2}(\pi a \omega r_{m})+\cos^{2}\frac{\pi}{2}\sqrt{1-4v_{m}a^2r_{m}^2}}$$
for $4v_{m}a^2r_{m}^2<1$.
\begin{table}[htb]
\caption{Critical values of angular momentum at $Q_{N}=.6$ are obtained by using $v_{f}$ at $r=2M$. The values of $l_{c}^{WKB}$ are present outside the bracket while inside is for $l_{c}^{QM}$.}
\label{tbl-smp1}
\vspace{2ex}
\begin{center}
\begin{tabular}{|c|c|c|c|}
\hline
\hline
 $\omega r_{m}$   & $s=0$    & $s=1$ & $s=2$\\
  \hline
  .5 &         1.283[1.352]          &       1.5[1.561]     &     2.042[2.090]\\
  \newline
  1 &         3.396[3.928]           &       3.5[3.531]     &     3.797[3.826]  \\
  1.5&         5.931[5.952]           &       5.5[5.521]   &     5.702[5.722]  \\
  2&         7.948[7.964]           &       7.5[7.515]      &     7.652[7.668]  \\
  \hline
  \hline

\end{tabular}
\end{center}
\end{table}
\begin{figure}[htb]
\begin{tabular}{cc}
\includegraphics[angle=0, width=0.46\textwidth]{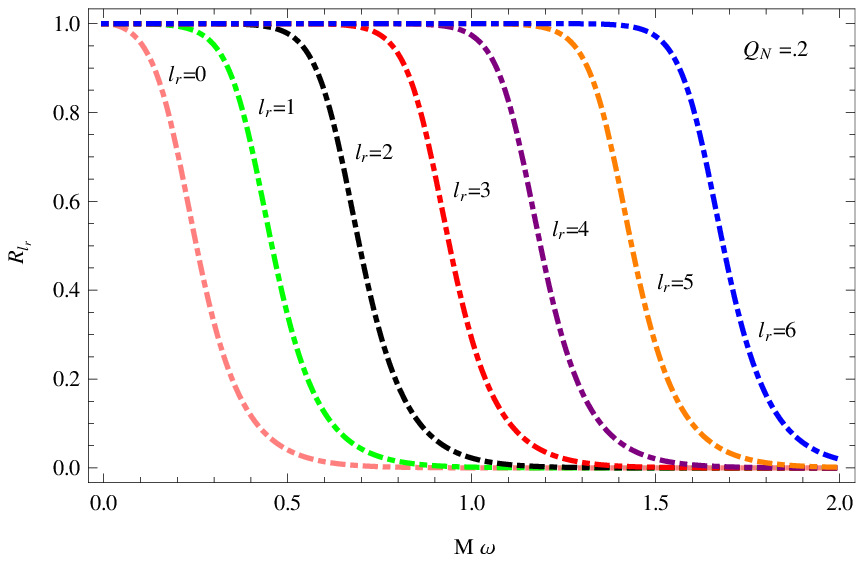}&
\includegraphics[angle=0, width=0.45\textwidth]{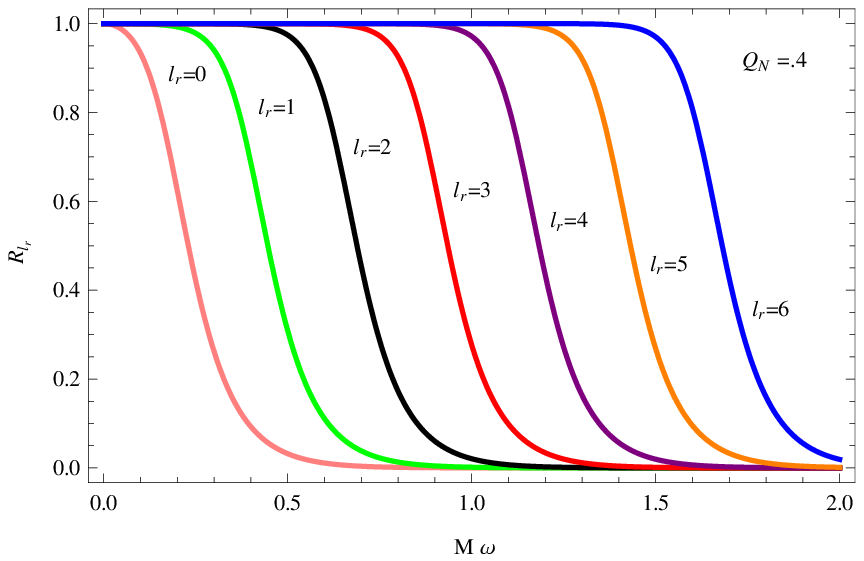}\\
\includegraphics[angle=0, width=0.46\textwidth]{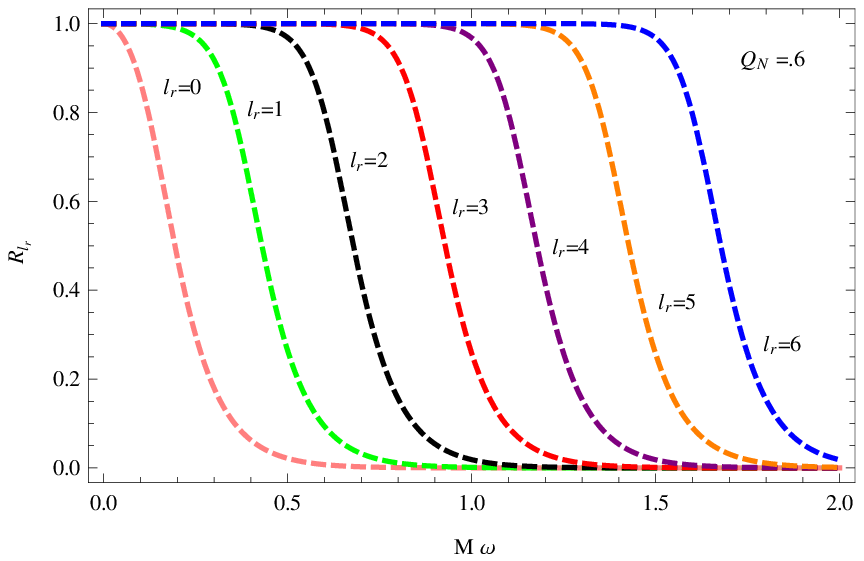}
\end{tabular}
\caption {Reflection coefficients for different choices of angular momentum for Reissner-Nordstrom black hole for spin $0$ particles.}
\label{fig-smp3}
\end{figure}

In fig.3 reflection coefficients are plotted for different values of angular momentum as a function of energy for $s=0$ case at $Q_{N}=.2,.4,.6$. Here we observed that $R_{l_{r}}$ tends to rapidly zero as $\omega\rightarrow\infty$. With the increasing charge reflection coefficient goes to zero earlier.
\begin{figure}[htb]
 \centering
  \includegraphics[width=10cm]{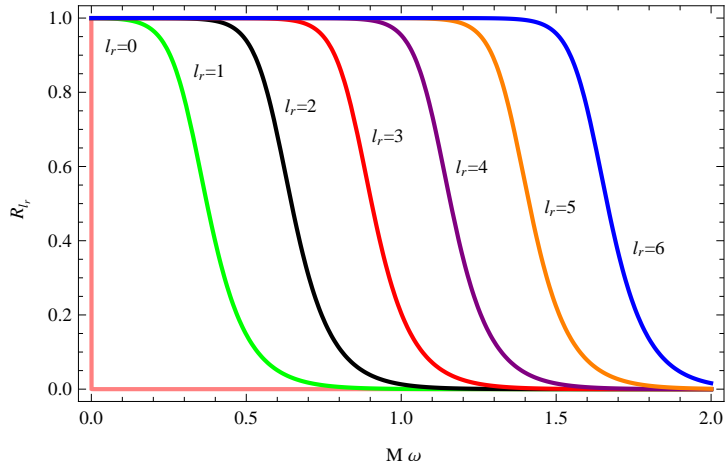}\\
  \caption{Reflection coefficients for different choices of angular momentum for Reissner-Nordstrom black hole for spin $1$ particles.}
  \label{fig-smp4}
\end{figure}
Fig.4 represents the reflection coefficient for scattering of electromagnetic wave from Reissner-Nordstrom black hole as a function of energy. An unusual behavior is noticed for $l_{r}=0$ in spin 1 case where the reflection coefficient got maximum value instantly and after that zero value is noticed with increasing of energy, this satisfies the condition that $l_{r}\geq s$ while for rest of the value of $l_{r}$ similar behavior is observed as for spin zero. Here term that includes charge is not present in spin 1 case so we don't have different choices for charge.
\begin{figure}[htb]
\begin{tabular}{cc}
\includegraphics[angle=0, width=0.46\textwidth]{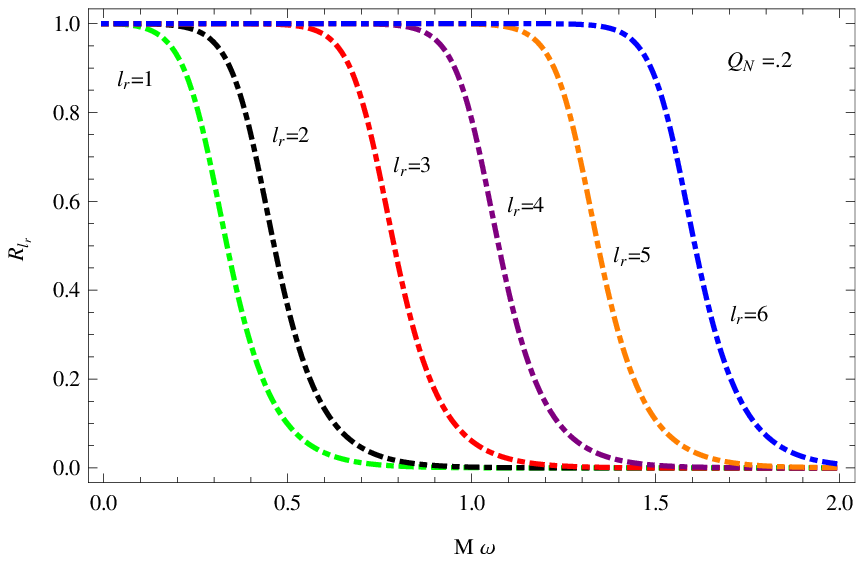}&
\includegraphics[angle=0, width=0.45\textwidth]{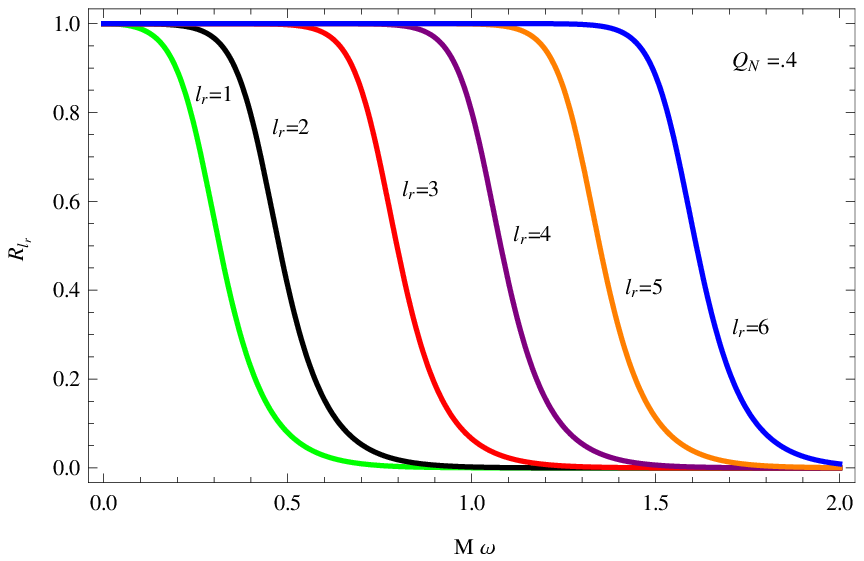}\\
\includegraphics[angle=0, width=0.46\textwidth]{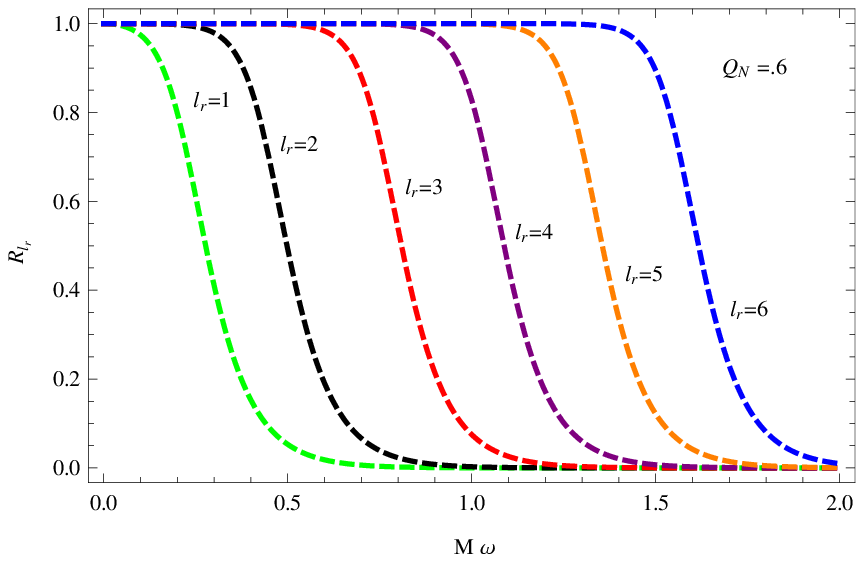}
 \end{tabular}
 \caption {Reflection coefficients for different choices of angular momentum for Reissner-Nordstrom black hole for spin $2$ particles.}
 \label{fig-smp5}
\end{figure}
Fig.5 depicts the reflection coefficient for the gravitational waves as a function of energy for the charged black hole. At the lower value of charge, difference between reflection coefficient for two consecutive lower values of angular momentum is small but it increased with the increasing of charge.
\begin{figure}[htb]
 \centering
  \includegraphics[width=10cm]{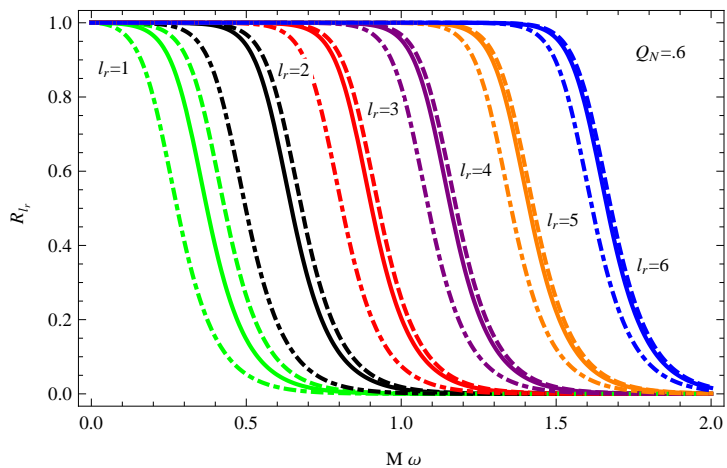}\\
  \caption{Comparison of reflection coefficients for different choices of angular momentum and spin.}
  \label{fig-smp6}
\end{figure}
In fig.6, we compare the reflection coefficients for the scattering of $s=0,1,2$ particles from Reissner-Nordstrom black hole at $Q_{N}=.6$. Dashed dotted line represents spin $0$ and solid line for spin $1$ while dashed line for spin $2$ particles. For the lower value of $l_{r}$ reflection coefficient differ with each other very clearly specially for the case $s=0$ and $1$ whereas for larger value of $l_{r}$ overlapping can be observed.
\textbf{\Large{\\[.2ex]Absorption Cross Section}}\\
Quantum mechanically total absorption cross section can be written as
$$\sum_{l_{r}=0}^{\infty}\sigma_{ab}^{(l_{r})}$$
where $\sigma_{ab}^{(l_{r})}$ denotes the partial absorption cross section and is,
$$\sigma_{ab}^{(l_{r})}=\frac{\pi}{\omega^2}(2l_{r}+1)(1-e^{2\iota\delta_{l_{r}}})$$
In the above expression $e^{2\iota\delta_{l_{r}}}$ represents phase shifts and defined by
$$e^{2\iota\delta_{l_{r}}=(-1)^{l_{r}+1}}  \frac{B_{\omega l_{r}}^{out}}{B_{\omega l_{r}}^{in}}$$
One can also obtain
\begin{equation}
\sigma_{ab}=\frac{\pi}{\omega^2}\sum_{l_{r}=0}^{\infty}(2l_{r}+1)|T_{l_{r}}|^2
\end{equation}
where $T_{l_{r}}$ is transmission coefficients. The most familiar form of eqn.6 is $$|T_{l_{r}}|^2+|R_{l_{r}}|^2=1$$
\begin{figure}[htb]
\begin{tabular}{cc}
\includegraphics[angle=0, width=0.46\textwidth]{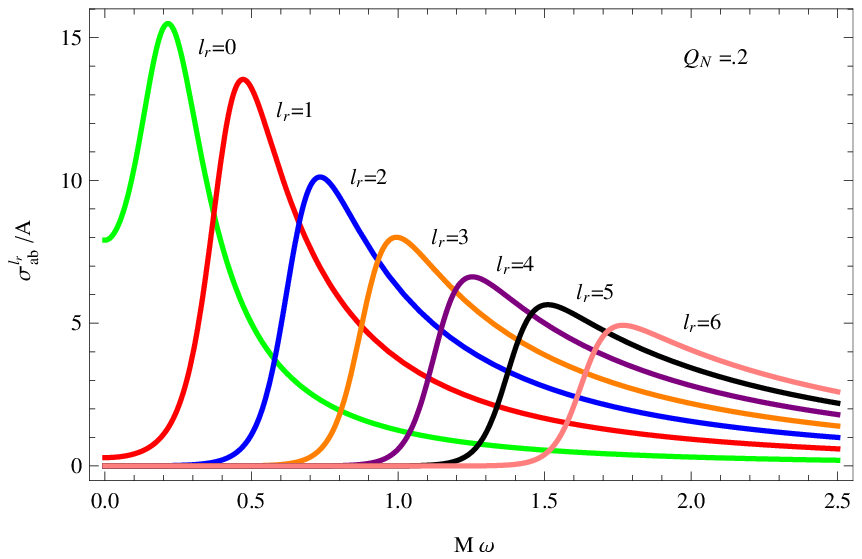}&
\includegraphics[angle=0, width=0.45\textwidth]{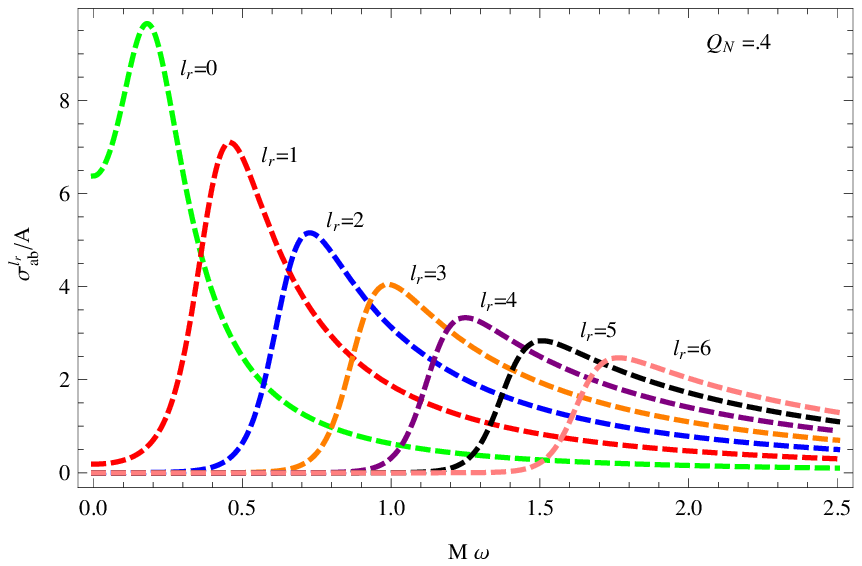}\\
\includegraphics[angle=0, width=0.46\textwidth]{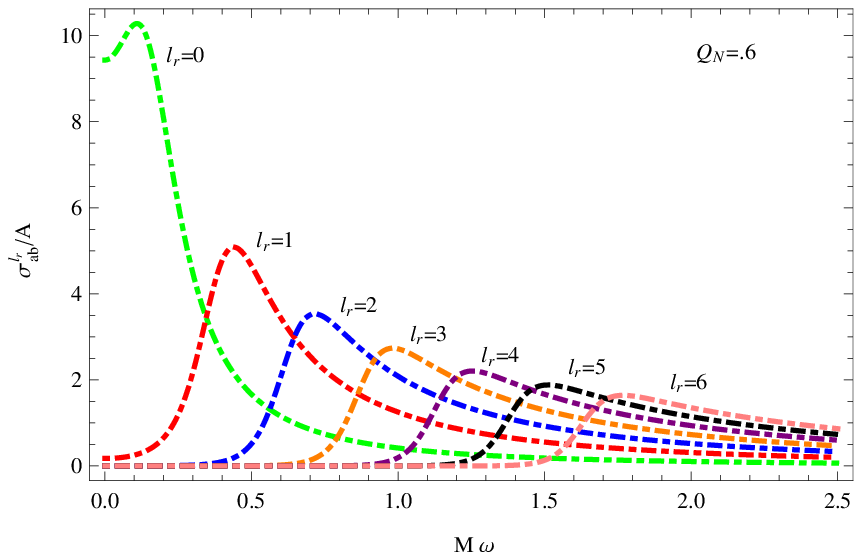}
 \end{tabular}
 \caption {Absorption cross section for different values of $l_{r}$ and $Q_{N}$ for spin zero case.}
\label{fig-smp7}
\end{figure}
Partial absorption cross section with $s=0$ and $Q_{N}=.2,.4,.6$ for different value of angular momentum is plotted in fig.7,only for $l_{r}=0$ non-vanishing cross section is observed in the limit of zero-energy whereas for remaining every value of $l_{r}>0$ starting value of corresponding absorption cross section zero is noticed,got maximum and then decreases asymptotically. For small value of $Q_{N}$ the cross section has greater value.
\begin{figure}[htb]
\begin{tabular}{cc}
\includegraphics[angle=0, width=0.46\textwidth]{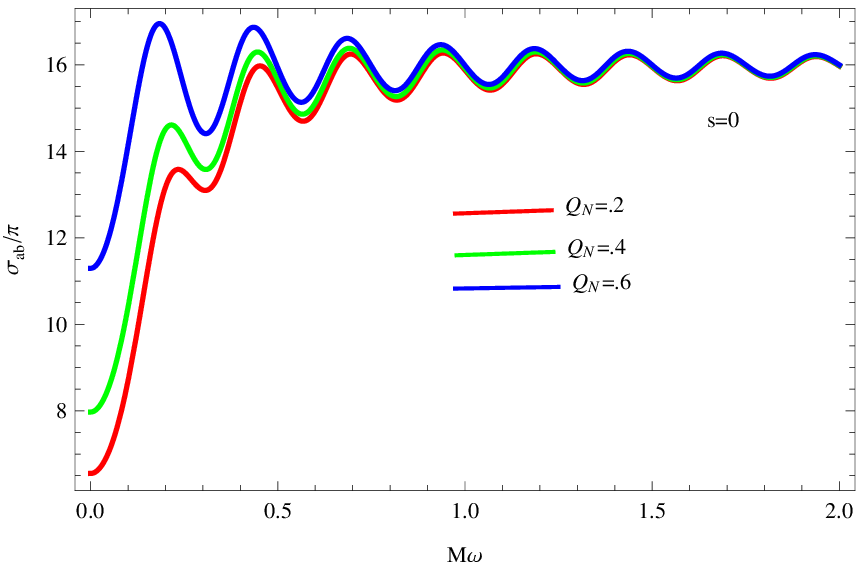}&
\includegraphics[angle=0, width=0.45\textwidth]{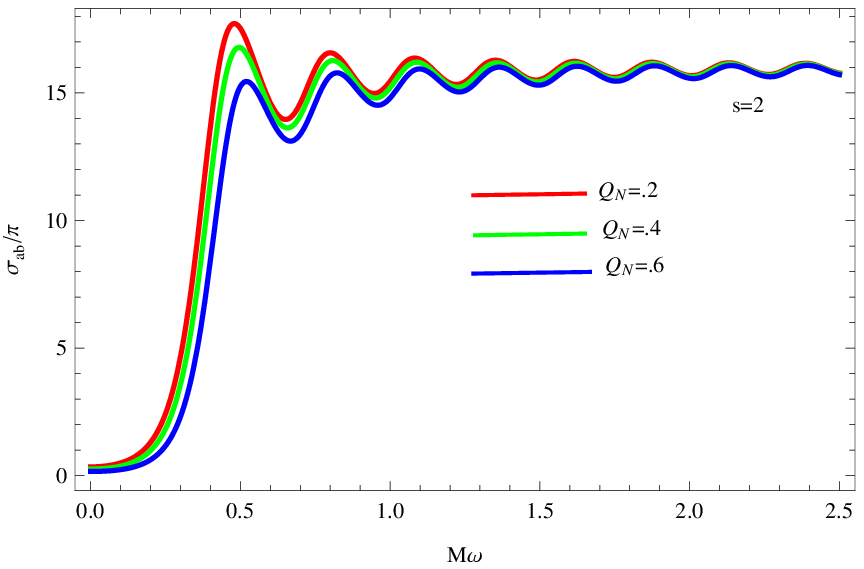}
 \end{tabular}
 \caption {Total absorption cross section for three different values $Q_{N}$ for the scalar and gravitational waves.}
 \label{fig-smp8}
\end{figure}
In fig.8,we present the total absorption cross section of extreme charged black hole for the case spin $0$ and $2$ at three different value of charge(i.e$Q_{N}=.2,.4,.6$)and summation for $l_{r}$ is performed up to $20$ in eqn.7. For small value of energy their behaviour is totally different from each other while for greater value i.e $M\omega>1$ almost similar pattern is observed. In case of spin $0$ as charge increases absorption cross section also increased in low energy regime but it is opposite in case of spin $2$ although the difference between the absorption cross section is very small for $Q_{N}=(.2,.4)and(.4,.6)$ for the gravitational waves.
\begin{figure}[htb]
\centering
\includegraphics[width=10cm]{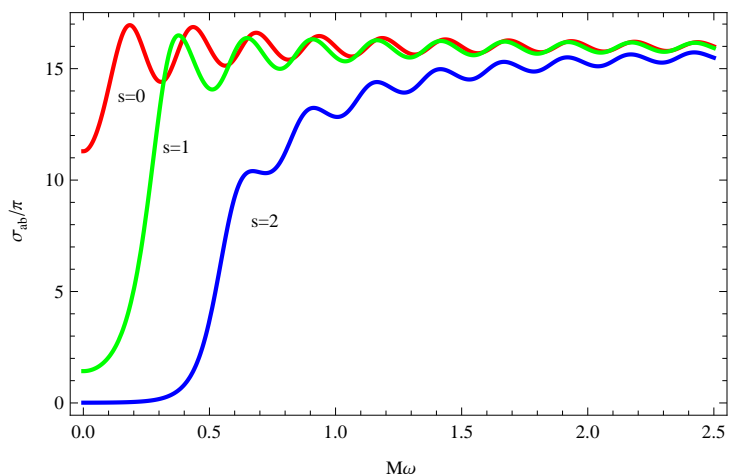}\\
  \caption{Comparison of total absorption cross section for three spins as a function of $M\omega$.}
  \label{fig-smp9}
\end{figure}
 A comparison is shown of total absorption cross section for scalar,electromagnetic and gravitational waves of the extreme Reissner-Nordstrom black hole at $Q_{N}=.6$ and take the summation over $l_{r}$ up to $20$. When $M\omega>>1$ regular oscillatory pattern is observed for the total absorption cross section and these oscillate around the classical range. It is also noted that the corresponding partial absorption cross section is represented by each local maximum.
 \textbf{\Large{\\[.2ex]Scattering Cross Section}}\\
 In Quantum Mechanics,well known form of scattering amplitude is expressed as
 $$g(\theta)=\frac{1}{2\iota \omega}\sum_{l_{r}=0}^{\infty}[e^{2 \iota\delta_{l_{r}}}-1]P_{l_{r}}(\cos\theta)$$
 By using this relation, we can give immediately differential scattering cross section
 $$\frac{d\sigma}{d\Omega}=|g(\theta)|^2$$
 We can also define scattering cross section\cite{26,27}
 \begin{equation}
 \sigma_{sc}(\omega)=\int\frac{d\sigma}{d\Omega}d\Omega=\frac{\pi}{\omega^2}\sum_{l_{r}=0}^\infty(2l_{r}+1)|e^{2 \iota\delta_{l_{r}}}-1|^2
 \end{equation}
\begin{figure}[htb]
\begin{tabular}{cc}
\includegraphics[angle=0, width=0.46\textwidth]{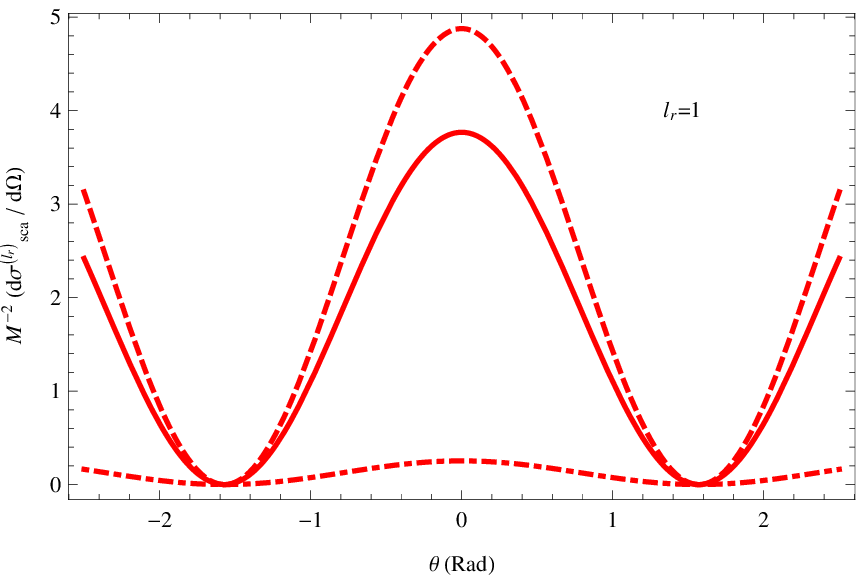}&
\includegraphics[angle=0, width=0.45\textwidth]{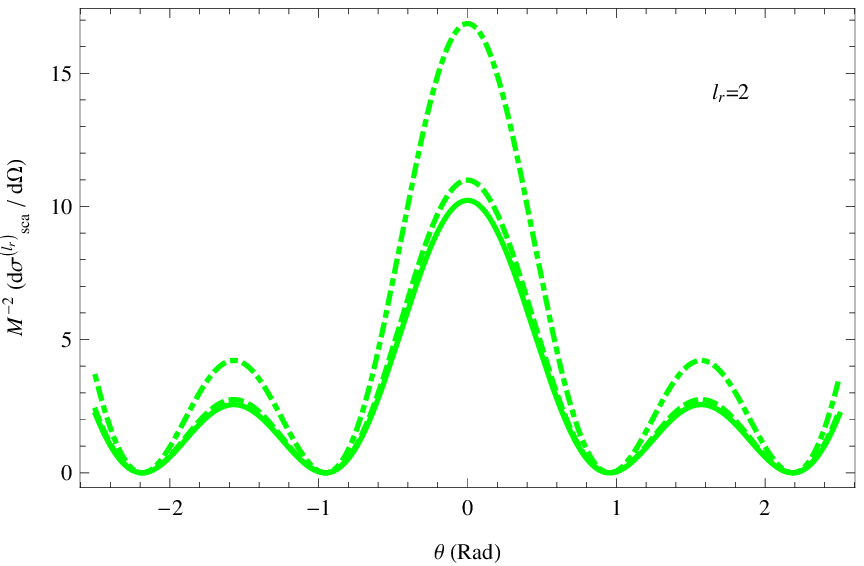}\\
\includegraphics[angle=0, width=0.46\textwidth]{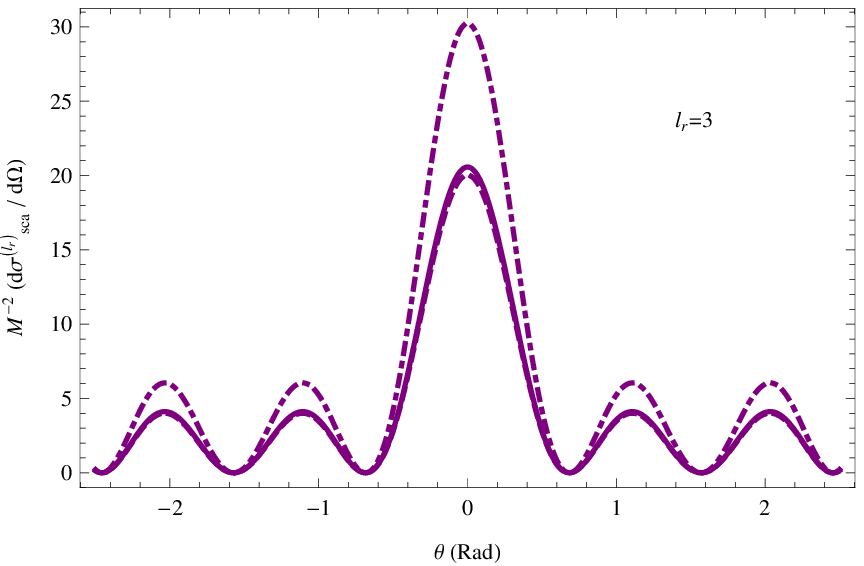}&
\includegraphics[angle=0, width=0.46\textwidth]{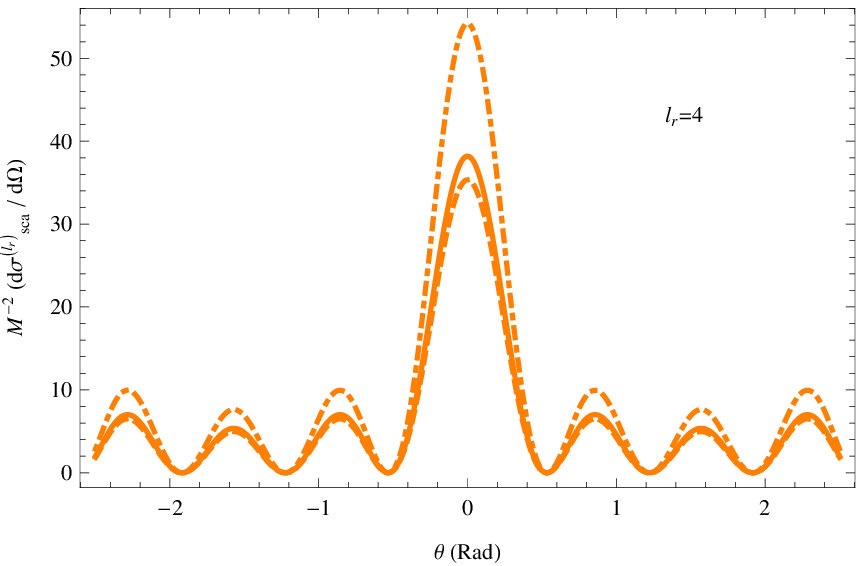}\\
\includegraphics[angle=0, width=0.45\textwidth]{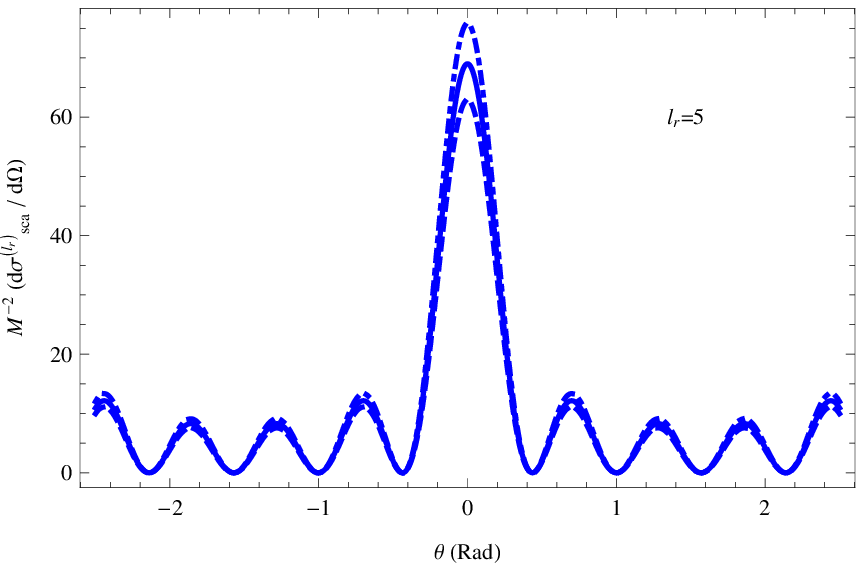}&
\includegraphics[angle=0, width=0.46\textwidth]{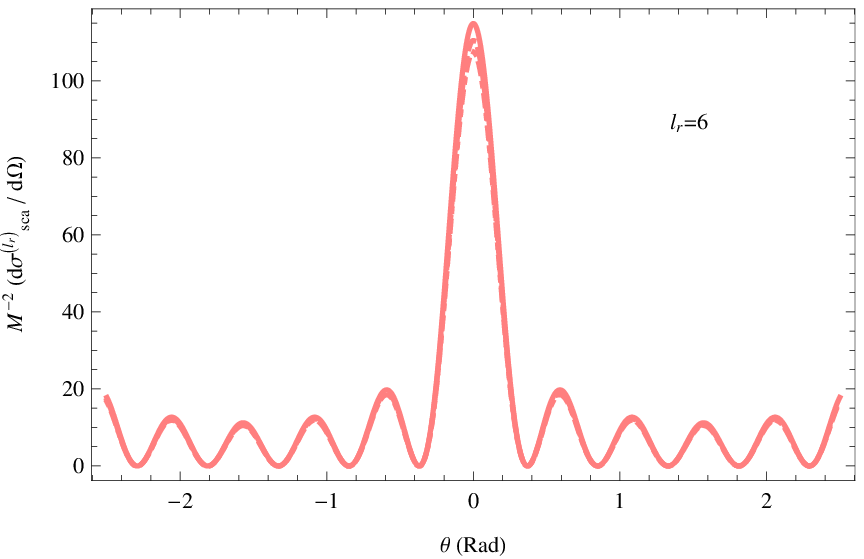}
 \end{tabular}
 \caption {Comparison of partial scattering cross section for three spins $0,1,2$ and different values of $l_{r}$ at $Q_{N}=.6$.  }
 \label{fig-smp10}
\end{figure}
Fig. 10 presents the comparison of partial scattering cross section for different values of angular momentum for spin 0,1 and 2 cases here scattering angle is in radian. For the lower values of $l_{r}$ greater difference of partial scattering cross section is noticed for three spins. As we increase the values of angular momentum difference becomes smaller and smaller while for $l_{r}=6$ overlapping can be seen.
\begin{figure}[htb]
\begin{tabular}{cc}
\includegraphics[angle=0, width=0.46\textwidth]{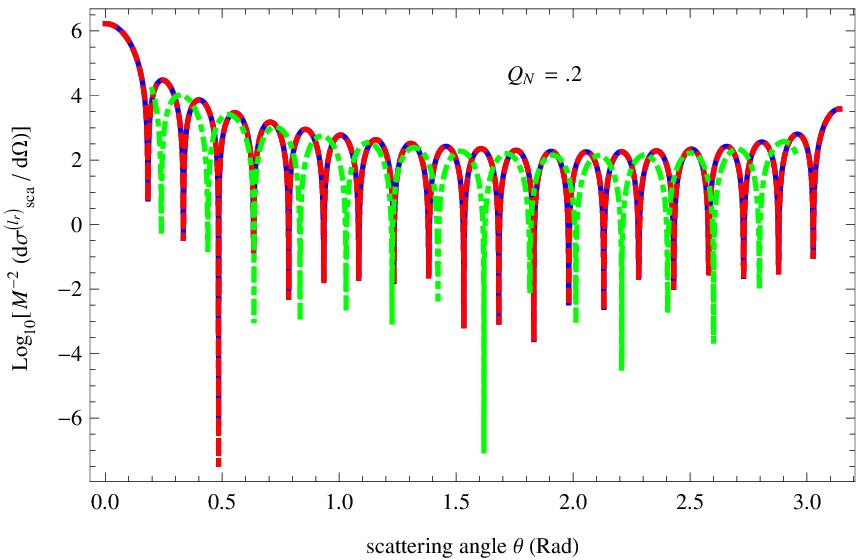}&
\includegraphics[angle=0, width=0.45\textwidth]{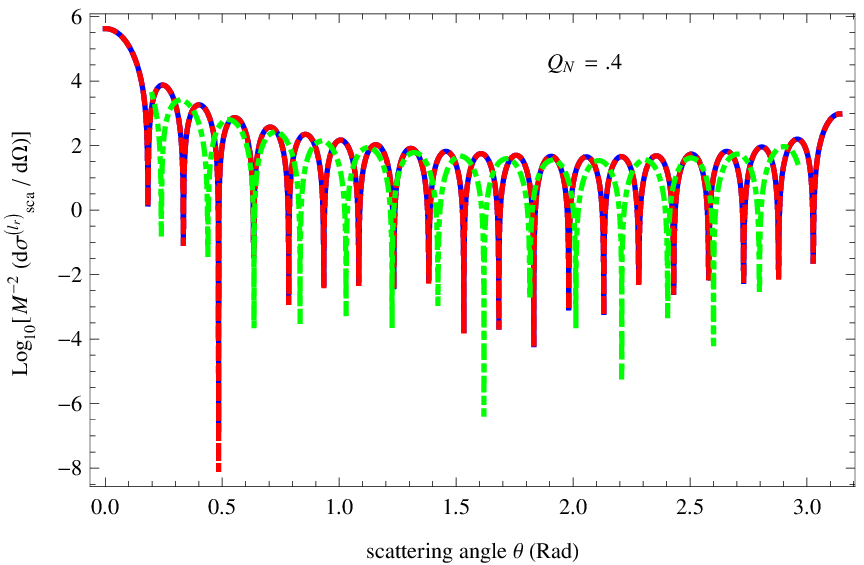}\\
\includegraphics[angle=0, width=0.46\textwidth]{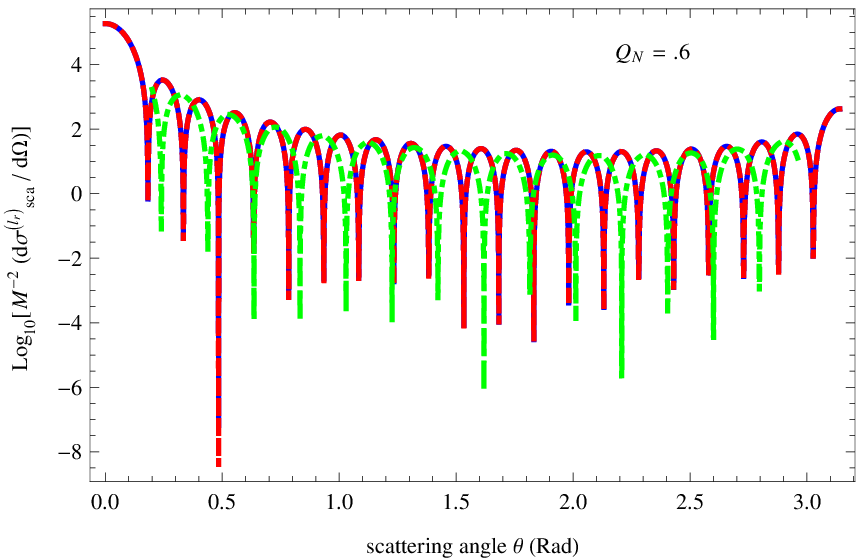}
 \end{tabular}
 \caption {Comparison of total scattering cross section for three spins $0,1,2$ and different values of charge}
\label{FIG}
\end{figure}
Fig.11 shows the scattering cross section for different values of charge and for scalar, electromagnetic and gravitational wave. Blue color is for spin $0$ and red is for spin $1$ while green represents spin $2$. For all three values of charge strengthened scattering flux is noticed in forward as well as in backward directions while its width becomes narrower. That's mean we can see glory phenomenon. For smaller value of charge scattering flux is more stronger as compered to larger value of charge.
\\
\textbf{\Large{\\[.2ex]Final Remarks}}\\
Let us summarize our results. We have presented an investigation of reflection and transmission coefficients for massless scalar, electromagnetic and gravitational waves as well as how these are affected by the black hole charge. In case of spin 0, reflection coefficients approaches to zero at initial values of energy as we increase the value of charge. For the smaller value of angular momentum we have shown same results for three different values of charge i.e. $.2, .4$ and $.6$. At the same value of charge, difference between the reflection coefficients are significant for the lower value of angular momentum while overlapping can be seen as the value of $l_{r}$ increases. Furthermore, it is also observed that critical values of angular momentum are linked with unstable orbit at $r=2M$ for spin $1$.
\newline
Moreover the height of the partial absorption cross section decreases with the increasing of charge. We also come to know that only for $l_{r}=0$,significant non-vanishing cross section is noticed in zero-energy limit in case of spin $0$. At a certain fixed value of charge, the total absorption cross sections for the three cases behave differently at lower values of energy. However, at larger values of energy, the regular oscillatory pattern observed is nearly similar. In the case of gravitational wave the starting value of total absorption cross section is zero whereas in the case of electromagnetic and scalar waves the starting value is non-zero.
\newline
Towards the end of the paper we calculate the scattering cross sections. We can see that with the increase of $l_{r}$, more complex oscillatory pattern is observed and differential scattering cross section has greater values but the width of the main scattering angle becomes narrower. The maximum of partial flux occurs at  $\theta=0$ in all the three cases. At constant charge, smaller values of angular momentum result in substantial difference between scattering cross section whilst overlapping is noted at large values of angular momentum.
\newline
Scattering flux is strengthened in both backward and forward directions. Glory phenomenon can also be observed. Scattering flux has greater value for lower value of charge while its value is decreasing with the increasing of charge. In future, more accurate results will be calculated by using numerical methods which include also spin part for complex black holes.


\begin{thebibliography}{99}
\bibitem{16}
 Eq. (2) in Ref. [11], which is the same as Eq. (2.6) in
Ref. [18], corresponds, indeed, to the equation of motion
of a massless vector field which does not couple to the
electric field of the black hole.
\end{thebibliography}

\end{document}